\documentclass[reprint, amsmath,amssymb, aps ]{revtex4-2}

\usepackage{graphicx}
\usepackage{dcolumn}
\usepackage{bm}
\usepackage{xcolor}
\usepackage{physics}
\usepackage{makecell}
\usepackage{mathtools}
\usepackage[normalem]{ulem}
\usepackage{tikz,xcolor,hyperref}

\definecolor{lime}{HTML}{A6CE39}
\DeclareRobustCommand{\orcidicon}{
	\begin{tikzpicture}
		\draw[lime, fill=lime] (0,0) 
		circle [radius=0.16] 
		node[white] {{\fontfamily{qag}\selectfont \tiny ID}};
		\draw[white, fill=white] (-0.0625,0.095) 
		circle [radius=0.007];
	\end{tikzpicture}
	\hspace{-2mm}
}

\foreach \x in {A, ..., Z}{\expandafter\xdef\csname orcid\x\endcsname{\noexpand\href{https://orcid.org/\csname orcidauthor\x\endcsname}
		{\noexpand\orcidicon}}
}

\definecolor{linkcolour}{HTML}{000066}	
\hypersetup{colorlinks,breaklinks,
	urlcolor=linkcolour, 
	linkcolor=linkcolour,
	citecolor=linkcolour}

\newcommand{\txtpow}[1]{{\mbox{\scriptsize{#1}}}}
\begin{document}
	
	\title{Label-free detection of single nanoparticles with disordered nanoisland surface plasmon sensor}
	
	\author{Hongki Lee$^{1,4}$}
	\author{Joel Berk$^{2,4}${\orcidA{}}}
	\author{Aaron Webster$^3$}
	\email{aaron.webster@canyon.k12.ca.us}
	\author{Donghyun Kim$^1${\orcidB{}}}
	\email{kimd@yonsei.ac.kr}
	\author{Matthew R. Foreman$^{2}${\orcidC{}}}%
	\email{matthew.foreman@imperial.ac.uk}
	\affiliation{$^1$School of Electrical and Electronic Engineering, Yonsei University, Seoul 03722, Korea}
	\affiliation{$^2$Blackett Laboratory, Imperial College London, Prince Consort Road, London, SW7 2BW, United Kingdom}
	\affiliation{$^3$Independent scholar, 187 Pinehurst Rd, Canyon, CA 94516, USA}
	\affiliation{$^4$Authors contributed equally}
	
	\date{\today}
	
	\begin{abstract}
 We report sensing of single nanoparticles using disordered metallic nanoisland substrates supporting surface plasmon polaritons (SPPs). Speckle patterns arising from leakage radiation of elastically scattered SPPs provides a unique fingerprint of the scattering microstructure at the sensor surface. Experimental measurements of the speckle decorrelation are presented and shown to enable detection of sorption of individual gold nanoparticles and polystyrene beads. Our approach is verified through bright-field and fluorescence imaging of particles adhering to the nanoisland substrate.
	\end{abstract}
	
	
	\maketitle

\section{Introduction}
 Recent years have seen great progress in the development of micro- and nanoscale optical technologies for biodetection. Sensitive and label-free detection of biomolecules such as viruses, DNA and proteins, is particularly important for implementing next-generation clinical assays and improved healthcare. Sensing platforms leveraging  (opto)-mechanical resonances \cite{Naik2009TowardsSpectrometry,LiOuLeiLiu2021}, whispering gallery modes \cite{KimBaaskeReview2017} and photonic crystals \cite{Cunningham2016SensorsJ} have seen great success in this field (see Table~\ref{tab:label_free_summary}). Surface plasmon resonance (SPR) sensors, i.e. those based on coupled electronic-optical oscillations at planar metal interfaces, are however currently  one of the leading commercial technologies  as they allow label-free multiplexed sensing, operation in aqueous environments and easy integration with microfluidic technology \cite{Homola2008,GuoPFBG2017}.  Despite their proven record, traditional SPR sensors rely on detecting resonance perturbations, e.g. frequency shifts, caused by analyte induced bulk refractive index changes and are thus currently not capable of single molecule measurements. Single molecule sensitivity is however desirable \cite{Taylor2017,CunninghamReview2020} since it facilitates rapid detection and allows non-equilibrium studies to be performed.

In the push to achieve greater sensitivity on an SPR platform, numerous strategies have been explored, including analyte labelling \cite{Shalabney2011}, signal amplification \cite{Zhou2018}, use of alternative materials such as aluminium or graphene, \cite{Knight2014,  Patil2019} and grating based designs \cite{Vempati2015,ZhuPS2020}. 
Motivated by the single molecule sensitivity achievable using localised surface plasmon resonances (LSPRs) \cite{Zijlstra2012OpticalNanorod,Acimovic2018AntibodyantigenBiosensors,Baaskearxiv}, the use of plasmonic assemblies and nano-structured substrates  has also received significant attention \cite{Duan2021}. Single plasmonic nanopores have, for example, enabled single molecule detection and spectroscopy \cite{Li2019DetectionNanopore,Freedman2016NanoporeTrapping,Brown2016NanoporeNanopore}, whilst nanogap and nanohole arrays have facilitated monitoring of biomolecular binding \cite{bios11040123,OH2014401,Cetin2015}. Surface immobilised nanoparticles \cite{ZijlstraReviewSmall}, nanoislands \cite{Chung2019}, nanowells and nanotubes \cite{Ye2014,McPhillips2010} have also been shown to enhance LSPR sensing. Alternatively, plasmon interferometry achieves greater sensitivity by leveraging interference of fields scattered by an analyte and a reference signal. As a result, plasmonic interferometers can sensitively monitor growth of thin protein layers, detect single exosomes and identify particle characteristics from either spatial interference or spectral fringes \cite{Yang2018,Qian2019,ZhangNatMethod2020}. 
More recently, developmental techniques based on magneto-plasmonics \cite{Mejia-Salazar2018} and quantum plasmon states for noise suppression \cite{Duan2021,Dowran2018} have also been investigated.

In this context this study presents a novel sensor capable of detecting adsorption of single nanoparticles using a traditional Kretschmann attenuated total reflection surface plasmon platform. By simultaneously exploiting sensitivity gains afforded by randomly nanostructured substrates, enhanced near fields and interferometric detection, we are able to overcome the limitations imposed by metallic losses and the resulting low $Q$ resonances. Specifically, by way of proof of principle we observe discrete sorption events for 50~nm radius gold nanoparticles and polystyrene beads, verified through optical imaging.

\begin{table*}[t!]
	\footnotesize
	\centering
	\begin{tabular}{c||c|c|c|c}
		Method&\makecell{Limit of \\detection (LOD)}&\makecell{Time \\resolution }&Sensing volume&\makecell{Commercially \\available}\\
		\hline\hline
		SPR  & 144 virons/mL  \cite{Takemura2021SurfaceTechnology,Chang2018SimpleAntibody}
		& $\sim10$~ns  \cite{Wang2017FastC,Zhou2020SurfaceAccess}
		& $\sim100\mbox{~nm}\times25~\mu\mbox{m}^2$
		&Yes \cite{Homola2008}\\
		\hline
		LSPR&  \makecell{1 molecule\\
			$\sim50$~kDa protein \cite{Zijlstra2012OpticalNanorod}\\
			$\sim10$~kDa protein \cite{Acimovic2018AntibodyantigenBiosensors}}
		&$\sim10$~ns \cite{Baaskearxiv}
		&$\sim10$~nm$\times 100\mbox{~nm}^2
		$&No\\
		\hline
		\makecell{Plasmonic \\ interferometric \\ scattering}&\makecell{1 molecule\\
			$\sim 150$~kDa protein \cite{ZhangNatMethod2020}}  
		& $\sim 50$~ms \cite{ZhangNatMethod2020}
		&$\sim100\mbox{~nm}\times500~\mu\mbox{m}^2$ \cite{ZhangNatMethod2020}
		&No\\
		\hline	
		Nanopore&\makecell{1 molecule\\
			$\sim1$~kDa DNA strand \cite{Li2019DetectionNanopore}}
			& $\sim0.01$~ms \cite{Li2019DetectionNanopore}
			&$\sim 4~\mu\mbox{m}\times100\mbox{~nm}^2$ \cite{Freedman2016NanoporeTrapping}
			&Yes \cite{Brown2016NanoporeNanopore}\\
		\hline
		\makecell{Photonic \\resonance}& \makecell{1 molecule\\
			$\sim2$~kDa DNA strand \cite{Baaske2014Single-moleculePlatform}\\
			$\sim10$~kDa DNA strand \cite{Ghosh2021ADiagnostics}}
		&$\sim1~\mu$s  \cite{Shu2015,DayanRingUp}
		&\makecell{$\sim 200\mbox{~nm}\times2000~\mu\mbox{m}^2$ (WGM)  \cite{PhysRevA.74.051802}\\$\sim 200\mbox{~nm}\times10000\mbox{~nm}^2$ (PC)  \cite{Wang2018MaximizingCavities}}
		&No\\
		\hline
		\makecell{Mechanical\\ resonance}&
		    \makecell{1 molecule\\
		    $\sim0.13$~kDa C$_{10}$H$_{8}$ \cite{Chaste2012AResolution}\\
			$\sim0.5$~kDa DNA base \cite{Jiang2020Mech}}
		&$\sim$10~ms \cite{Chaste2012AResolution}
		& \makecell{ $\sim$500$~\mu\mbox{m}^2 $* (cantilever) \cite{Waggoner2007Micro-Detection}\\
			$\sim 500$~nm$^2$* (nanotube) \cite{Chaste2012AResolution}
}	
		&No\\
		\hline 
		\makecell{Nanowire\\ conductance}&  \makecell{1 molecule\\
			$\sim0.5$~kDa DNA base \cite{Sorgenfrei2011Label-freeTransistor,Li2020Single-MoleculeScience}}
		& $\sim1$~ms \cite{Zhou2017AdvancesBioelectronics} &\makecell{$\sim1000\mbox{~nm}^2$*  \cite{Zafar2018SiliconCharacteristics}}
		&No
		
	\end{tabular}
	\caption{Summary of performance of leading label free biosensing methods. The LOD refers to either the limit of detection in analyte concentration, or analyte particle size for systems with single molecule sensitivity. Sensing volume is given for a single sensing element and is expressed as a distance from sensor surface multiplied by an effective sensing area as dictated by the relevant mode distribution/structure dimensions. Multiplexed arrays can increase the effective sensing volume. * indicates analyte must be tightly attached to surface.}
	\label{tab:label_free_summary}
\end{table*}

\section{Detection principle}

A traditional Kretschmann-type surface plasmon sensor comprises of a thin metallic film deposited on a glass substrate, with a lower refractive index  medium, typically an aqueous solution containing analyte particles, on the opposing side. Light is obliquely incident from the glass substrate at an angle so as to optimally excite surface plasmon polaritons (SPPs) in the film. Destructive interference between the specularly reflected incident light and the field re-radiated by SPPs excited in the film then produces a notch in the reflected angular spectrum \cite{Raether1988}. The location of the reflection dip is sensitive to local refractive index variations in the analyte medium, thus enabling detection of bulk analyte adsorption onto the metallic sensor surface \cite{Homola2008}. SPPs can however undergo scattering from surface features or roughness as they propagate. One possible fate of such scattered SPPs is to couple into $s$- and $p$-polarised waves propagating away from the interface, which produces 
a weak diffuse far-field scattering pattern observable in both the analyte medium and substrate \cite{Novotny2006}. This diffuse scattering has recently been used to image single proteins in an interferometric scattering microscopy (iSCAT) type setup \cite{ZhangNatMethod2020,LiNatCom21}. Alternatively, SPPs can couple into in-plane SPPs before undergoing absorption, further scattering or re-radiation back into the glass substrate. Conservation of momentum in SPP-to-SPP scattering, however means that re-radiated SPPs are confined in the far field to a thin annular cone of `leakage radiation' with an opening angle and angular width approximately equal to that of the reflection dip \cite{Raether1988,Simon1976a}. In principle, the leakage cone could thus be used for bulk refractive index sensing, however since it exhibits the same angular sensitivity as the attenuated total reflection notch, and the limit of detection is equally dictated by material absorption losses, such an approach affords little advantage. Critically, the leakage cone however also possesses strong intensity fluctuations around the ring. This one-dimensional manifestation of optical speckle
arises from interference of randomly scattered SPPs on the
sensor surface \cite{Schumann2009} and represents a unique fingerprint of the underlying scattering microstructure. As such the random speckle is highly sensitive to the motion and configuration of individual scatterers near the sensor surface \cite{Berkovits1991,Berkovits1990}, especially in the multiple scattering regime \cite{Berk2021} where material absorption can play an important role \cite{Berk2021a}. 

Adsorption of a nanoparticle, such as a virus or protein, to the sensor surface modifies the scattering microstructure, in turn producing a change in the leakage ring speckle intensity which can be observed. Specifically, the leakage intensity $I(\mathbf{r})$ after particle adsorption is given by
\begin{align}
I(\mathbf{r}) &\sim \epsilon^{1/2}(\mathbf{r})[|\mathbf{E}_{\txtpow{ref}}(\mathbf{r})|^2 + |\delta\mathbf{E}(\mathbf{r})|^2  \nonumber\\ &\quad\quad+2|\mathbf{E}_{\txtpow{ref}}(\mathbf{r})||\delta\mathbf{E}(\mathbf{r})|\cos\Phi(\mathbf{r})], \label{eq:I}
\end{align} 
where $\epsilon(\mathbf{r})$ is the electric permittivity distribution, $\mathbf{E}_{\txtpow{ref}}$ is the random (reference) field when the analyte particle is absent, $\delta\mathbf{E}$ is the change in the scattered field upon introduction of the analyte particle and $\Phi$ describes the relative phase and polarisation of $\mathbf{E}_{\txtpow{ref}}$ and $\delta\mathbf{E}$. In the renormalised Born approximation for a small spherical particle at position $\mathbf{r}'$ the change in the field is, to leading order, given by \cite{Chaumet1998,Foreman2017}
\begin{equation}
	\delta\mathbf{E}(\mathbf{r}) \approx 4\pi  \epsilon_h k_0^2 a^3 \frac{(\epsilon_p - \epsilon_h)}{\epsilon_p + 2 \epsilon_h} G(\mathbf{r},\mathbf{r}') \mathbf{E}_{\txtpow{ref}}(\mathbf{r}') ,\label{eq:Es}
\end{equation}
where $k_0$ is the vacuum wavenumber of light, $a$ and $\epsilon_p$ are the radius and electric permittivity of the analyte particle, $\epsilon_h = \epsilon(\mathbf{r}')$  is the background permittivity of the host medium and $G(\mathbf{r},\mathbf{r}')$ represents the system Greens function \cite{Tomas1995}. As per physical intuition we thus see that the scattered field is equivalent to that from a dipole with moment $\mathbf{p} = \alpha \mathbf{E}_{\txtpow{ref}}$, where $\alpha = 4\pi a^3 \epsilon_h (\epsilon_p - \epsilon_h) /(\epsilon_p + 2 \epsilon_h)$ is the polarisability of the analyte particle \cite{Bohren1983a}. Through an appropriate choice of polarisability, resonances in the analyte particle (as can be present in for example plasmonic nanoparticles or quantum dots) can be described.  Notably, for small bioparticles scattering such resonances are not relevant and $\alpha$ is small. Consequently, $|\delta\mathbf{E}|\ll |\mathbf{E}_{\txtpow{ref}}|$ which in turn implies that the interference term in Eq.~\eqref{eq:I} is dominant.

\begin{figure}[t]
	\centering
	\includegraphics[width=\columnwidth]{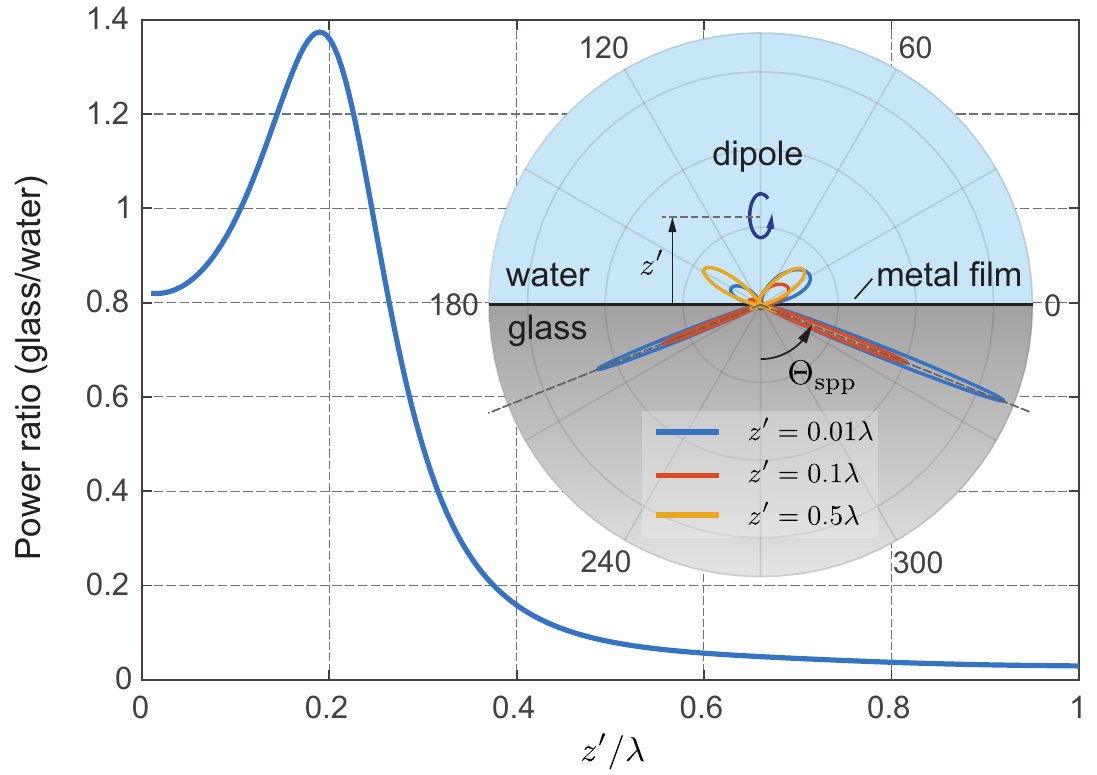}
	\caption{Ratio of the total power scattered into the glass substrate and aqueous superstrate from a dipole scatterer as a function of height above the metallic film structure described in the main text. Polar inset depicts a cross-section through the radiation pattern for dipole scatterers at heights of $0.01\lambda$ (blue), $0.1\lambda$ (orange) and $0.5\lambda$ (yellow). Excitation wavelength was taken at $\lambda = 721$~nm. 
		\label{fig:greens}}
\end{figure}

The reference field $\mathbf{E}_{\txtpow{ref}}$, as discussed above, arises from random scattering of SPPs on the surface, and thus can be represented using an order of scattering expansion \cite{TsangKong}, such that  $\mathbf{E}_{\txtpow{ref}} = \mathbf{E}_{\txtpow{spp}} + \mathbf{E}^{(1)} + \cdots$, where $ \mathbf{E}_{\txtpow{spp}}$ is the SPP field excited in a planar film and $ \mathbf{E}^{(j)}$ are scattering corrections ($j=1,2,\ldots$). Similarly, $G(\mathbf{r},\mathbf{r}')$ is the system Greens tensor accounting for surface disorder and can thus be expressed as a Dyson series \cite{TsangKong}
\begin{align}
G(\mathbf{r},\mathbf{r}') &= G_0(\mathbf{r},\mathbf{r}') 
\nonumber\\
&\quad+\int G_0(\mathbf{r},\mathbf{r}'')V(\mathbf{r}'') G_0(\mathbf{r}'',\mathbf{r}') d\mathbf{r}''+ \cdots ,
\end{align}
where $G_0(\mathbf{r},\mathbf{r}')$ is the Greens function for a smooth planar film structure. Note that the approximations made here imply loop type scattering trajectories are neglected, however,  these can be accounted for through use of a dressed particle polarisability. The leading order change in the field $\delta \mathbf{E}^{(1)}(\mathbf{r}) \approx G_0(\mathbf{r},\mathbf{r}') \mathbf{E}_{\txtpow{spp}}(\mathbf{r}')$  exhibits the same strong directional scattering into the glass substrate as the reference field  when the analyte particle is close to the surface. 
 Figure~\ref{fig:greens}, for example, depicts the far-field angular radiation pattern $\sim|G_0|^2$, for a dipole scatterer with unit moment, in water, at distances of $0.01\lambda$, $0.1\lambda$ and $0.5\lambda$ from a metallic film composed of a 50~nm thick gold and and 2~nm thick chromium layer on glass. Calculations assumed refractive indices of $1.33$, $0.13+4.24i$, $3.06+ 3.40i$ and $1.51$ respectively and a free space wavelength of $\lambda = 721~$nm \cite{Johnson1972}. As per Eq.~\eqref{eq:Es}, the scatterer dipole moment was assumed to be aligned with the exciting SPP field $\mathbf{E}_{\txtpow{SPP}} \sim (1i,0,-3.21)$ \cite{Maier2007} producing an asymmetric radiation pattern \cite{Kretschmann1972,Mueller2013}, albeit in this case the asymmetry is weak since the $z$-component of $\mathbf{E}_{\txtpow{spp}}$ dominates. From Figure~\ref{fig:greens} it is evident that as the analyte particle moves further from the surface, diffuse scattering into the analyte (water) medium becomes dominant over the more tightly confined scattering into the glass substrate. To quantify this  effect, the ratio of the power scattered into the glass and water half-spaces is also shown in Figure~\ref{fig:greens} as a function of particle distance $z'$ from the metallic film. Whilst for small distances the power ratio is approximately unity, a significant drop-off is seen for distances greater than approximately $\lambda/4$. Physically, the decrease in the power scattered into the substrate arises from the exponential decay of the SPP away from the sensor substrate (the out-of-plane intensity decay length of the SPP into the water solution is calculated to be $\sim 130$~nm). At larger distances this decay means that light scattered from the analyte particle couples weakly into SPPs, such that coupling to more diffuse propagating waves is the only allowed scattering pathway. Higher order scattering terms do not change the directional scattering, but instead serve to introduce speckle fluctuations in $\delta \mathbf{E}$.

In light of these scattering characteristics, two distinct advantages to sensing via detection of the leakage speckle can thus be identified. Firstly, the sensing volume is more strongly confined to the surface of the sensor than if sensing were performed using diffuse scattering into the analyte medium. Heuristically, the field scattered into the analyte medium will fall as $\exp[- \kappa_z z']$, where $\kappa_z$ is the out-of-plane component of the SPP wavevector, due to the decreasing illumination strength intrinsic to $\mathbf{E}_{\txtpow{spp}}$. On the other hand, the field scattered into the leakage ring decreases by virtue of the decay in $\mathbf{E}_{\txtpow{spp}}$ and additionally due to the $\sim \exp[-\kappa_z z']$ drop off of the Greens function (cf. Figure~\ref{fig:greens}). Secondly, since  scattering into the leakage ring exhibits greater angular confinement than scattering into the upper half-space, obtainable signal to noise ratios are higher. In particular, for the configuration considered in Figure~\ref{fig:greens}, the peak intensity is at best approximately 8 times higher in the leakage ring, than in the diffuse scattering pattern.
 
\section{Sensitivity analysis}

To quantify changes in the leakage speckle over time as individual particles adhere to the sensor surface, we propose use of the Pearson correlation coefficient, $C$, between a reference background speckle pattern $I_{\txtpow{ref}}$ (taken at time $t=0$) and the observed speckle $I$ at a later time $t=\tau$, defined as 
\begin{equation}
	C = \frac{\mbox{cov}[I_{\txtpow{ref}},I]}{\sigma_{\txtpow{ref}}\,\sigma}  \label{eq:PCC}
\end{equation} 
where the speckle patterns are sampled at positions $j = 1,\ldots,N$ in the ring,  $\sigma_{(\txtpow{ref})}^2 = \sum_{j=1}^N [I_{(\txtpow{ref}),j}-\mu_{(\txtpow{ref})}]^2 / (N-1)$, $\mbox{cov}[I_{\txtpow{ref}},I] = \sum_{j=1}^N[I_{\txtpow{ref},j}-\mu_{\txtpow{ref}}][I_j-\mu] / (N-1)$ and $\mu_{(\txtpow{ref})}= \sum_{j=1}^N I_{(\txtpow{ref}),j} / N$ are the spatial variance, covariance and average of the intensity around the ring respectively. Use of $C$ means changes in the speckle structure can be measured whilst removing the effect of any spurious global intensity fluctuations \cite{Webster2018}.

To gain some insight into the sensitivity afforded through use of the speckle correlation we can estimate the approximate expected change in $C$ upon nanoparticle adsorption using a random phasor sum based approach. Assuming the intensity distribution $I_{\txtpow{ref},j}$ describes a well developed speckle (whereby $\langle I_{\txtpow{ref},j} \rangle = \langle (I_{\txtpow{ref},j} - \langle {I}_{\txtpow{ref},j} \rangle )^2 \rangle^{1/2}$ where $\langle \cdots \rangle$ denotes an average over the ensemble of possible speckle patterns or equivalently surface scattering configurations) and that analyte particle scattering is weak, it follows that after particle adsorption $I_j$ also describes a well developed speckle. Depending on the properties of the surface roughness, $I_{\txtpow{ref},j}$ can exhibit a slowly varying envelope function in addition to the rapid speckle variations \cite{Raether1988}. If, however, the number of sample positions in the speckle ring, $N$, is large and the average speckle size is much smaller than the variations in any envelope function we can assume piecewise ergodicity holds, whereby $\mu_{(\txtpow{ref})} = \overline{\langle {I}_{(\txtpow{ref}),j} \rangle}= \overline{\langle ({I}_{(\txtpow{ref}),j} - \langle {I}_{(\txtpow{ref}),j} \rangle)^2 \rangle }^{1/2} = \sigma_{(\txtpow{ref})}$, where $\overline{g}$ denotes a spatial average of $g$ around the ring. For a \emph{fixed} background speckle realisation, the average change in $C$ after particle adsorption is thus approximately
\begin{equation}
	E[\Delta C ]\approx 1- \frac{N}{N-1}\left[\sum_{j=1}^N \frac{I_{\txtpow{ref},j} }{N\mu_b \mu}E[I_j]  - 1\right].
\end{equation}
where $E[\cdots]$ are expectations taken with respect to analyte particle binding position on the sensor surface. For weak particle scattering $\mu \approx \mu_{\txtpow{ref}}$ such that
\begin{equation}
	E[\Delta C ]\approx-\frac{1}{N \mu_{\txtpow{ref}}^2}  \sum_{j=1}^N  I_{\txtpow{ref},j}E[\Delta_j]
\end{equation}
where  we have let $I_j = I_{\txtpow{ref},j} + \Delta_j$ and assumed $N$ is large. 
$I_j$ derives from the sum of amplitudes $U_j = U_{\txtpow{ref},j} + \delta U_{j}$, where $\delta U_{j}$ is the field at position $j$ scattered from the analyte particle and $I_{\txtpow{ref},j} = |U_{\txtpow{ref},j}|^2$ whereby $\Delta_j = 2 \mbox{Re}[ U_{\txtpow{ref},j} \delta U_{j}^* ] + |\delta U_{j}|^2$. Note that in the leakage ring, the field is predominantly $p$-polarised such that a scalar approach is valid. 

The statistics of $\delta U_{j}$ depend on the statistics of the field incident upon the analyte particle $\mathbf{E}_{\txtpow{ref}}$.  We here consider two limiting cases, namely when the illuminating field is either a uniform plane SPP (i.e. $\mathbf{E}_{\txtpow{ref}} = \mathbf{E}_{\txtpow{spp}} $) or a fully developed Gaussian plasmonic speckle \cite{Foreman2019a}, which correspond to the single and multiple SPP scattering regimes respectively. In both cases we propagate the field from the random scatterer position to the observation position using the Greens function, $G_0(\mathbf{r}_j,\mathbf{r}')$ and 
 assume the detection position is far  from the sensor surface such that amplitude variations are negligible For the single scattering regime, and neglecting SPP absorption, it follows that $\delta U_{j}$ has a constant amplitude and a uniformly distributed phase. Noting from above that the dipole moment induced in the scatterer is dominated by the $z$ component, it follows that $|\delta U_{j}|$ is approximately uniform around the speckle ring implying $E[\Delta_j]$ is a constant, which we denote $I_{s}$. 

Alternatively, for the multiple scattering regime, $\delta U_{j}$ is modelled as a complex Gaussian circular random variable. Accordingly, $I_j$ obeys the modified Rician distribution \cite{Goodman2006}, with mean $\langle I_j \rangle = I_{\txtpow{ref},j} + E[\Delta_{j}]$ where $E[\Delta_{j}]$ is the average intensity scattered to position $j$ by a single analyte particle. Importantly, $E[\Delta_{j}] $ is again constant around the speckle ring ($\triangleq I_{s}$), since angular effects associated with orientation of the induced dipole average out due to the speckled nature of the local illuminating field. We thus ultimately find that in both cases the expected change in the correlation coefficient scales as
\begin{equation}
	E[\Delta C ]\sim-\frac{I_{s}}{\mu_{\txtpow{ref}}} \label{eq:EDC}
\end{equation}
albeit the precise value of $I_{s}$ varies with the degree of multiple scattering. The corresponding sensitivity and limit of detection are $S = |\partial E[\Delta C]/\partial I_{s} |\approx \mu_{\txtpow{ref}}^{-1}$ and $I_s^{\txtpow{LOD}} \approx m \sigma_C \mu_{\txtpow{ref}}$, where  $\sigma_C$ is the standard deviation of experimental noise on $C$ and $m$ dictates the detection threshold. It is important to  note that the approximate analysis given here only  considers multiple scattering present in the exciting field and thus neglects loop scattering paths and further scattering of waves after interaction with the analyte particle. As discussed in detail in Ref.~\cite{Berk2021} such scattering processes can play an important role in dictating device sensitivity. 

Equation~\eqref{eq:EDC}, although only an approximate rule of thumb, can be used to gain important insight into the sensor design. For example, use of thicker gold films will typically reduce both $I_s$ and $\mu_{\txtpow{ref}}$ equally and thereby has little effect on the average change in speckle correlation, however it does degrade both the sensitivity and detection limit of the system. Conversely, dipole scatterers typically scatter more strongly into higher index materials due to the associated increase in the density of states. $I_s$ and $\mu_{\txtpow{ref}}$ can therefore be increased through appropriate choice of substrate, bearing in mind this will also modify the SPP coupling conditions. Material composition of the analyte particle however solely affects $I_s$ (c.f. Eq.~\eqref{eq:Es}). Furthermore, if multiple ($M$) particles adsorb to the sensor, interference between each individual scattering component must be considered in the averaging process, such that $I_{s}$ scales sublinearly with respect to $M$. The speckle correlation referenced to a fixed frame will thus slowly decay to zero as multiple particles bind to the sensor surface, at a rate dependent on $I_s/\mu_{\txtpow{ref}}$. Nevertheless, the slow reduction in the average correlation change can be mitigated by intermittently resetting the frame which is taken as the reference. This also helps overcome the tradeoff between the sensitivity and range of detection (i.e. maximum number of detectable particles) inherent to most sensors, albeit ultimately detection will be limited by the increase in $\mu_{\txtpow{ref}}$ and thus the smaller average correlation change as compared to the noise $\sigma_C$. 

In our calculation of the mean expected correlation change given above, averaging over the possible particle adsorption position meant that interference terms appearing in $\Delta_j$ averaged to zero. The sensitivity predicted by Eq.~\eqref{eq:EDC}, whilst providing a guide to average performance, therefore does not necessarily capture the ability of our proposed sensor to observe a single binding event, which in particular exploits the SNR enhancement offered through optical interference. To quantify this, we can instead consider the standard deviation of the change in correlation upon particle adsorption which we find to be  $\sigma_{\Delta C} \approx  [12 I_{s}/\mu_{\txtpow{ref}}]^{-1/2}$. Noting then that field scattered by a single analyte particle scales as $I_{s} \sim a^6$ (c.f. Eq.~\eqref{eq:Es}), the typical change in $C$ upon particle adsorption scales as $a^3$, such that smaller particles can in principle be detected as compared to pure scattering based approaches. Further discussion of the noise  performance of the proposed sensor is given in \cite{Berk2020}.

\begin{figure*}[t]
	\centering
	\includegraphics[width=\textwidth]{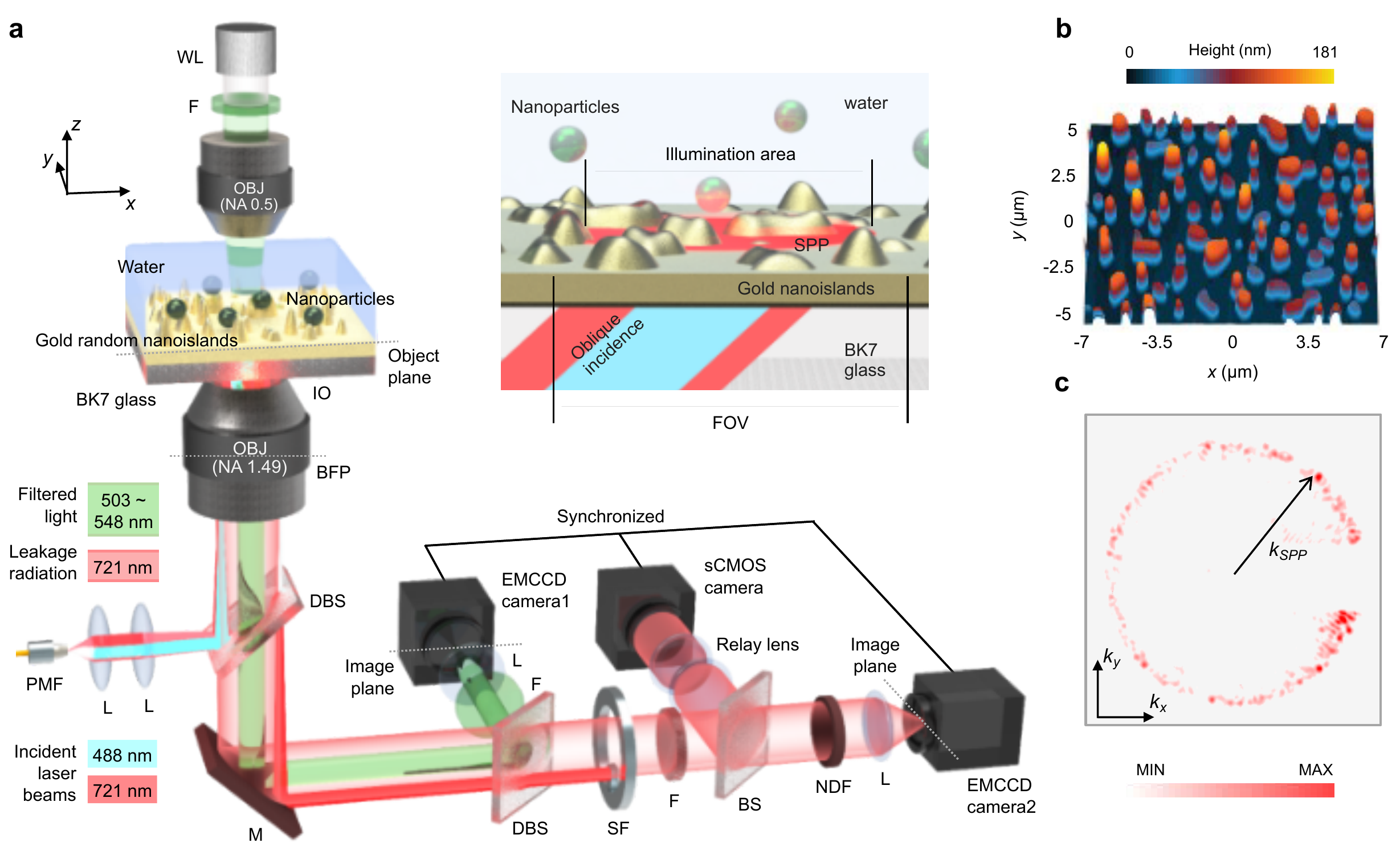}
	\caption{(a) Optical setup (WL: white light, F: filter, OBJ: objective, NA: numerical aperture, IO: immersion oil, BFP: back focal plane, DBS: dichroic beam splitter, L: lens, PMF: polarisation-maintaining fibre, M: Mirror, SF: spatial filter, BS: beam splitter, NDF: neutral-density filter, and FOV: field of view). (b) An atomic force microscopy image of a gold nanoisland substrate.  (c) A BFP image of speckled leakage ring radiation.}   
	\label{fig:exp_setup}
\end{figure*}

\section{Experimental method}
\subsection{Optical setup}

 To demonstrate the principle of our proposed sensing method we use the experimental setup shown in Figure~\ref{fig:exp_setup}(a). Excitation of SPPs was achieved using a $p$-polarised illumination laser source of wavelength $\lambda = 721$~nm (MRL-FN-721-30 mW, CNI, China) which was manipulated using a polarisation-maintaining fibre, lenses, an optical iris, a dichroic beam-splitter (DBS, T505lpxr, Chroma Technology Corporation, USA), and an objective (UAPON 100XOTIRF, $\mbox{NA} = 1.49$, Olympus, Japan),   
  so as to be incident at an angle of $\Theta_{\txtpow{spp}}=68.3^\circ$, corresponding to the SPP coupling angle. The laser power at the specimen was maintained at 6.37~W/cm$^2$.
    
  In order to promote SPP scattering and generation of speckled leakage radiation a gold film substrate with randomly distributed nanoislands was used. To fabricate the gold nanoisland film, a 15~nm thick gold film was prepared on a BK7 glass substrate by thermal deposition after cleaning with sonication in acetone, isopropyl alcohol, and distilled water. Random nanoislands were then formed by an annealing process on a hot plate at 550$^\circ$C for 4 hours. Additional thermal deposition was subsequently performed to produce a chrome adhesion layer and gold film with 2 and 50~nm thickness. The deposition rate of chrome and gold was set to be 0.2 and 0.7 \r{A}/s, respectively. The morphology of the fabricated gold nanoisland film, as found using an atomic force microscope (AFM), is shown in Figure~\ref{fig:exp_setup}(b). Depending on fabrication parameters, the geometrical properties of the random nanoislands follow normal distributions with different size and separations \cite{Yoo2021,Chen2015}. Further details of the synthesis and statistics of the nanoisland substrates can be found in Refs.~\cite{DonKimSmall2010,Fab1,Fab2}. From analysis of Figure~\ref{fig:exp_setup}(b), the size ($A$), height ($H$), and  nearest separation distance ($SD$) of the random nanoislands were determined to be $A = 0.15 \pm 0.074~\mu$m$^2$, $H = 83 \pm 16$~nm, and $SD = 1.0 \pm 0.2~\mu$m. Multiple nanoisland samples were fabricated and the reproducibility of the sensing layers was confirmed. Nanodot structures formed by sophisticated techniques, e.g., electron-beam lithography, could facilitate more predictable SPP scattering albeit at the expense of complexity and cost.
  
   Optical images of the random surface plasmon leakage radiation were observed in both the back-focal and image planes by first separating the scattered light using a beam splitter (BP145B1, Thorlabs, Inc., USA), before acquisition by a scientific CMOS (sCMOS, Zyla, Andor Technology Ltd., UK) camera and EMCCD camera 2 (imagEM, Hamamatsu Photonics K.K., Japan), respectively. A bandpass filter (ET720/60m, Chroma Technology Corporation, USA) and custom-built spatial filter, were used to ensure only surface plasmon leakage radiation was detected. A representative image of the resulting random surface plasmon leakage radiation ring is shown in Figure~\ref{fig:exp_setup}(c). The speckle contrast around the ring was found to be close to unity for all cases presented here indicating that the speckle patterns were fully developed \cite{Goodman2006}. The SPP wavelength $\lambda_{\txtpow{SPP}}$ and attenuation length can be extracted from the angle and width of the leakage ring \cite{Raether1988} and were found to be $514$~nm and $0.7~\mu$m respectively. Comparing to the theoretical values for the leaky SPP mode in a water-gold-chrome-glass planar stratified system of $515$~nm and $6.1~\mu$m respectively we note that the intensity attenuation length is significantly reduced due to surface scattering as would be expected, whilst the resonance wavelength is only slightly shifted. Calculations assumed the same planar structure as in Figure~\ref{fig:greens}. Using our experimental parameters we find that $k_{\txtpow{SPP}} l = 3.9 \times 10^3$, where $k_{\txtpow{SPP}} = 2\pi/\lambda_{\txtpow{SPP}}$ and $l = (n\sigma)^{-1} =  316~\mu$m is the mean free path evaluated using the experimental areal scatterer density $n$ and assuming the nanoisland scattering cross-section $\sigma$ is calculated as described in Ref.~\cite{Shchegrov1997}. Multiple scattering is typically considered strong when the Ioffe-Regel criterion, $k_{\txtpow{SPP}}l \ll 1$, is satisfied \cite{Berkovits1990}. Accordingly we see that SPP scattering on the nanoisland substrate is dominated by single scattering events. 
   
    Sensing experiments reported in this work used either 50~nm radius gold nanoparticles (GNPs, 753688-25ML, Sigma-Aldrich Corporation, USA) or  50~nm radius fluorescently labelled polystyrene nanobeads (F8803, Thermo Fisher Scientific Inc., USA) as the target analyte particles. The plasmonic resonance of the GNPs lay at $\sim 570$~nm and was thus not coincident with the illumination wavelength.  Fluorescent nanobeads were pumped using a $488$~nm laser source (35-LAP-431-230, CVI Melles Griot, USA) passing through the same optical chain as the $721$~nm source (see Figure~\ref{fig:exp_setup}). The excitation laser power was fixed at 25.46~W/cm$^2$ at the specimen. Nanoparticle suspensions were diluted with phosphate-buffered saline solution to a concentration of $\sim 3$~pM.
   
\begin{figure}[t]
    	\centering
    	\includegraphics[width=\columnwidth]{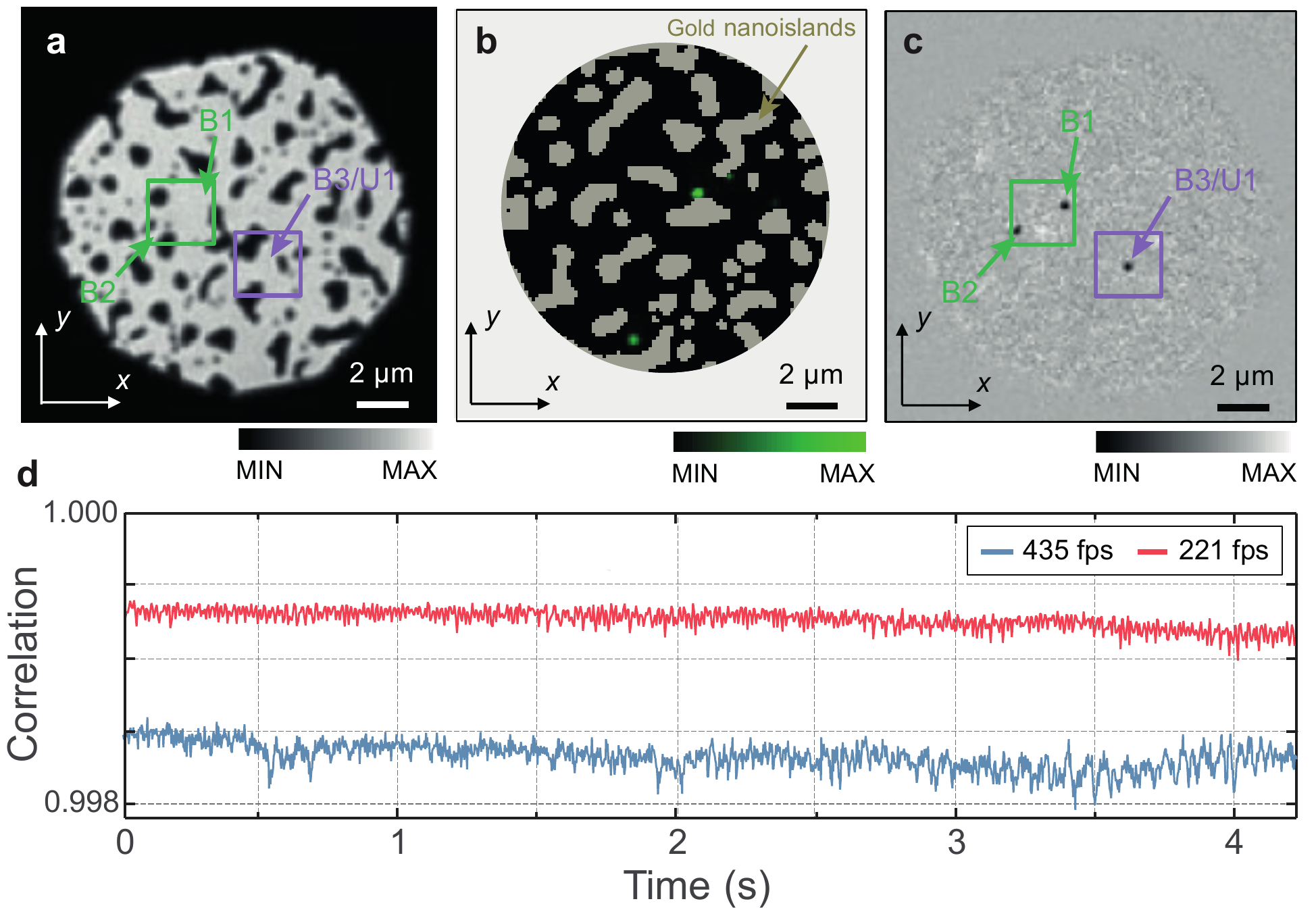}
    	\caption{(a) Typical bright field image of nanoisland substrate. Green and purple regions and labels correspond to data plotted in Figure~\ref{fig:gnp}.  (b) Example fluorescence image of nanobeads, overlaid on binary image of the nanoisland distribution as found from corresponding bright field image.  (c) Example cumulative binding image derived from bright field images in which three bound GNPs are evident. (d) Control curves depicting noise level on correlation coefficient before averaging was performed for frame rates of 435 (blue) and 221~fps (red). }   
    	\label{fig:images}
    \end{figure}

   To verify that changes in the leakage speckle correspond to binding or unbinding of analyte particles to the sensor surface, two additional imaging systems were also incorporated into our experimental design. GNPs scatter light strongly enough such that we could simultaneously observe them using a bright field microscope setup, as shown in Figure~\ref{fig:exp_setup}(a). Specifically, the gold nanoisland surface and gold nanoparticles were observed using an electron-multiplying charge-coupled device camera (EMCCD camera 1, iXON 897, Andor Technology Ltd., UK) through a long-pass dichroic beam splitter (DBS, DMLP550R, Thorlabs, Inc., USA) and a band-pass filter (BPF, FF01-525/45-25, Semrock, Inc., USA). A broadband halogen lamp (OSL2, Thorlabs, Inc., USA), a bandpass filter (FF01-525/45-25, Semrock, Inc., USA) and condensor lens (LMPlanFLN, $\mbox{NA} = 0.5$, Olympus, Japan) were used as the illumination source. An optical iris was adjusted such that the observation area was larger than the illumination area. An example bright field image of the nanoisland substrate and GNPs is shown in Figure~\ref{fig:images}(a). The scattering cross-section of polystyrene nanobeads is however significantly smaller than for GNPs due to the lower magnitude refractive index of $\sim 1.58$, such that analyte particles could not be easily observed using the bright field system. Instead, nanobeads were fluorescently labelled and fluorescence images were acquired using EMCCD1.  An example fluorescence image is shown in Figure~\ref{fig:images}(b), in which we also overlay a binarised bright field image showing the positions of nanoislands on the sensor surface.

    All three cameras were operated to capture optical images at the same frame rate with a synchronised trigger. The pixel size of EMCCD cameras 1 and 2 was 156 and 56~nm and images were acquired 5 hours after the laser sources were switched on so as to ensure stable output power. Cameras used acquisition speeds of 200 and 435~fps to obtain the dynamics of fluorescence nanobeads and GNPs, respectively. Note that in a sensing experiment only detection of the leakage speckle is necessary. In this work the additional imaging systems,  which would not easily provide sufficient image contrast to visualise smaller biological scatterers, are only used for corroboration of the sensing signal.

\subsection{Data processing}

\subsubsection{Leakage speckle}
To calculate $C$ from the experimental speckle images, acquired data were first averaged over 8 and 4 frames for sensing of GNPs and fluorescent nanobeads respectively, so as to mitigate the effects of noise (see Figure~\ref{fig:images}(d) for control curves without averaging). Noting the images were collected at frame rates of 435 and 200~fps, the corresponding time resolutions were $\sim 18$ and 20~ms. For each time-averaged speckle image, the centre and dimensions of the speckle ring were extracted by numerically minimising the Hilbert angle \cite{FranksBook} between the average image and a Gaussian profile ring-like windowing function using a simplex search method implemented by the Matlab {\tt{fminsearch}} function \cite{Lagarias1998}. Variations in the registration of the leakage ring and the camera were found to be minimal throughout the course of individual experimental runs, such that a fixed annular mask was used to extract only pixel values within the leakage ring, which were subsequently used to evaluate Eq.~\eqref{eq:PCC} for each frame. 
The Matlab-based step finding algorithm, {\tt{findchangepts}} was finally used to extract abrupt changes in the mean correlation \cite{Killick2012} which were ascribed to individual particle adsorption or desorption events.

\subsection{Bright-field and fluorescence images}
Bright field and fluorescence images were also processed digitally to identify discrete adsorption and desorption events to verify step changes observed in the Pearson correlation coefficient based sensorgram. Bright field and fluorescence images of GNPs and fluorescent nanobeads respectively were again averaged over 8 and 4 frames to reduce noise.
To remove the background surface features present in the bright field images the difference between subsequent time averaged frames was taken and their mean subtracted. A cumulative binding image was subsequently formed by integrating the differential images over time (see Figure~\ref{fig:images}(c) for a representative example).  Cumulative images were binarised using a threshold value derived from the baseline image noise. Distinct regions were then analysed to determine their total number, individual centroids and sizes. Single pixel regions were discarded and transient events were filtered by rejecting binarised regions which moved beyond their own extent over a few frames. Fluorescence images were subject to a similar processing chain, however since the nanoislands were not fluorescent and thus no background surface features were observed, binarisation could be performed directly on the fluorescence images.

\section{Results and discussion}

\begin{figure}[t]
	\centering
	\includegraphics[width=\columnwidth]{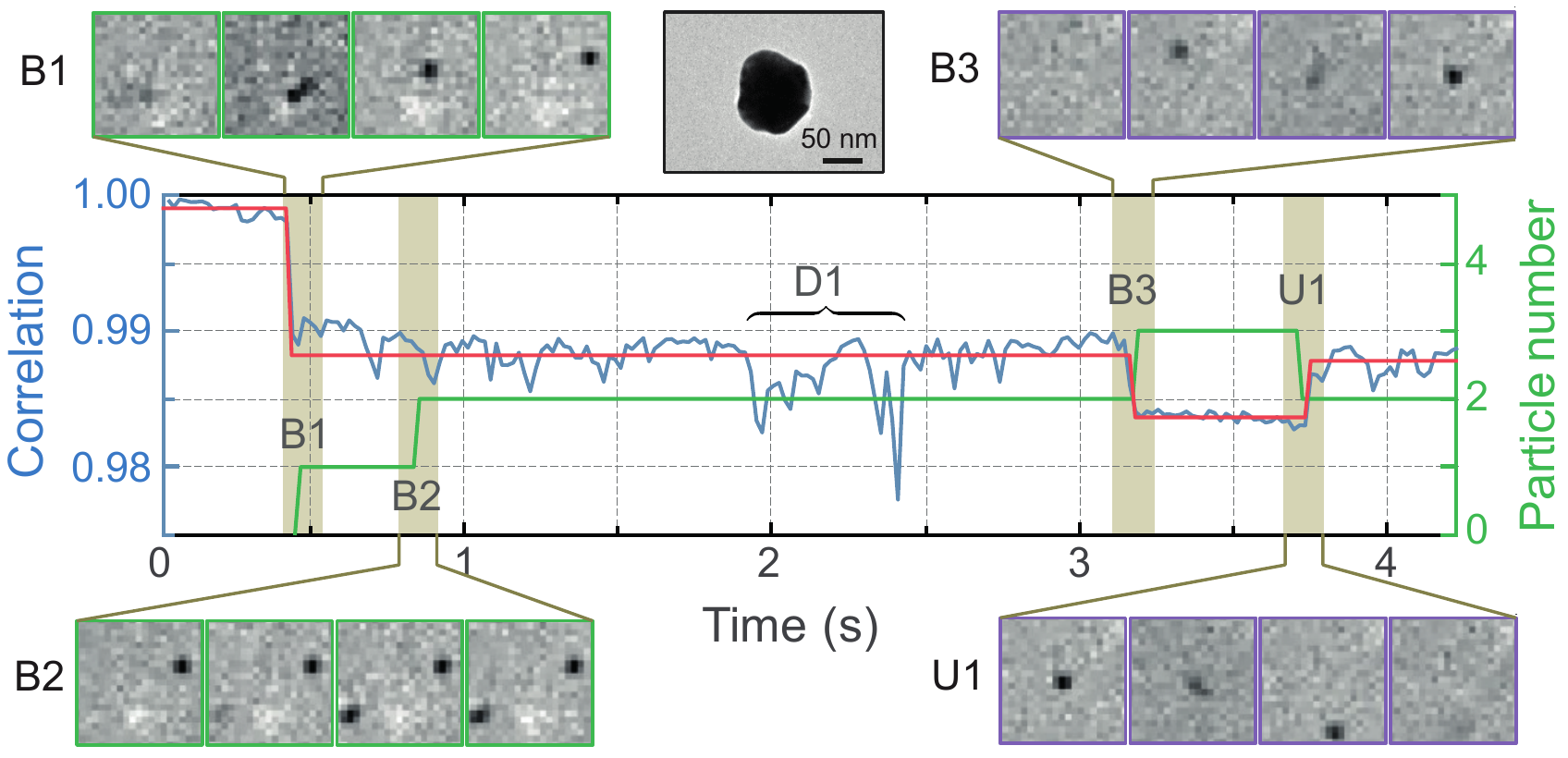}
	\caption{Example $C$ sensorgram (blue) obtained for sensing of 50~nm radius GNPs (see inset for a TEM image), with extracted steps  (orange) and particle count (green) found from analysis of bright field images.   Also shown are the sequence of images of regions highlighted in Figure~\ref{fig:images}(a) and (c) by green and purple squares for events B1, B2, B3 and U1 indicated. See also Supplementary Movie 1.}   
	\label{fig:gnp}
\end{figure}

As an initial demonstration, sensing of 50~nm radius GNPs was performed. A transmission electron microscope (TEM) image of a typical GNP is shown in the inset of Figure~\ref{fig:gnp}. A sample of the Pearson correlation sensorgram output, calculated from the final masked images as discussed above is plotted in blue in Figure~\ref{fig:gnp}. Output from the step finding algorithm is also shown by the solid orange line. A number of clear step-like events (labelled as B1, B3 and U1) are evident suggestive of individual particle adsorption and desorption. Note, throughout this work we assign each event a letter indicating the type of event (B, D or U corresponding to binding, volume diffusion, or unbinding respectively) and an incremental numeric index, e.g. B1 denotes the first binding/adsorption event in the corresponding sensorgram. Upon careful inspection of the bright field images for the frames preceding and following the steps (see Supplementary Movie 1), it is apparent that the B1, B3 and U1 events can be attributed to adsorption and desorption of nanoparticles at the positions indicated in Figure~\ref{fig:images}(a) and (c). Sequences of some key frames in the cumulative binding image within the regions of interest (also highlighted in Figure~\ref{fig:images}(a) and (c)) are also shown in the associated blow-outs. Image sequences corresponding to B1 and B3 show a GNP diffusing towards the sensor surface before adsorption confirming our interpretation of the sensorgram changes. Similarly the image sequence for event U1 shows that the GNP which initially bound during B3, desorps and diffuses away from the substrate. In addition to adsorption events, GNPs diffusing near the surface can give rise to stochastic transient events (e.g. D1 shown in Figure~\ref{fig:gnp}) in both the correlation sensorgram and bright field images. Such diffusion noise is less significant at lower analyte concentrations. Such transient events were filtered as part of the bright field processing algorithm. The resulting particle count, as indicated in green in Figure~\ref{fig:gnp}, exhibits a clear correspondence with the speckle  correlation coefficient. Analysis of the binding images, however, shows an additional event B2 which is not evident in $C$. The B2 image sequence shown in Figure~\ref{fig:gnp} shows the corresponding GNP binding. ``Non-detection'' events like B2 can occur due to the nature of the near field which is formed from scattering of the SPPs. Specifically, in certain positions the higher order scattering terms in the series expansion of $\mathbf{E}_{\txtpow{ref}}$ can produce positions at which $\mathbf{E}_{\txtpow{ref}}\approx \mathbf{0}$. If an analyte particle adsorps in a region in which the local field is weak, the resulting perturbation to the far field leakage speckle is correspondingly reduced \cite{Yang2014}.

A further feature evident from Figure \ref{fig:gnp} is that following a particle binding event, e.g. B1, the magnitude of fluctuations in $C$ increases. GNP adsorption in our experiments occurs primarily via physisorption, such that the binding strength is relatively weak. Correspondingly, adsorped GNPs can still undergo random motion on the nanometre scale within the trapping potential which produces the increase in sensorgram noise. Although this effect is noticeable for GNPs, it would be markedly reduced for more weakly scattering particles and for stronger binding processes e.g. binding of a protein to a receptor.

\begin{figure}[t]
\centering
\includegraphics[width=\columnwidth]{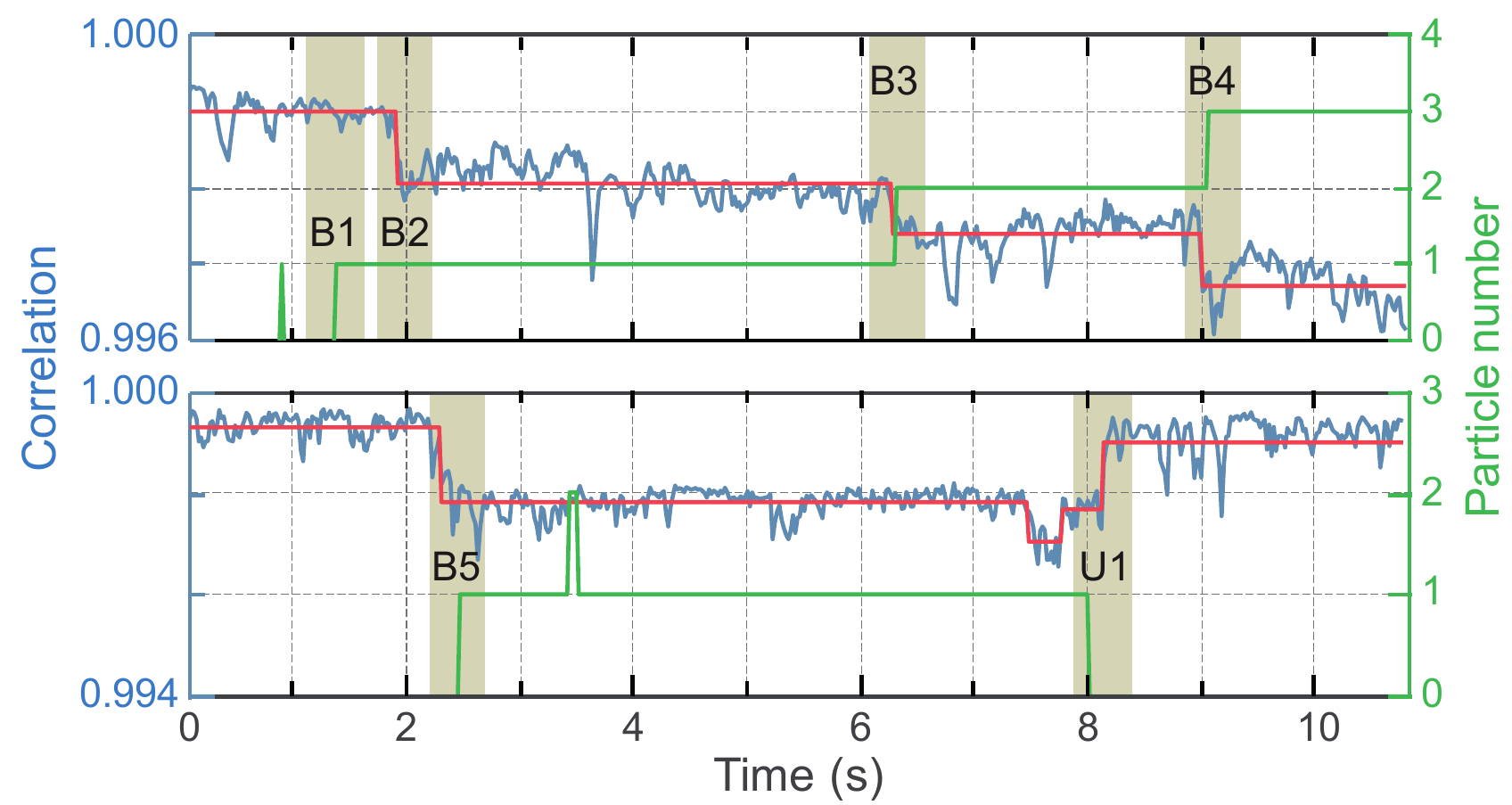}
\caption{Experimental $C$ sensorgrams (blue) exhibiting step-like features upon  sorption of 50~nm radius polystyrene nanobeads as identified using a step finding algorithm (orange). The time dependence of the particle count extracted from fluorescence images is also shown in green. See also Supplementary Movies 2 and 3. }   
\label{fig:nanobead}
\end{figure}

As a further demonstration of our proposed method, measurements of diffusion and adsorption of 50~nm radius polystyrene nanobeads were performed. Figure~\ref{fig:nanobead} shows two example sensorgrams with corresponding curves  from the step finding algorithm (orange) and particle count analysis (green, which is now performed on the raw fluorescence image). Notably, typical step sizes in the Pearson coefficient sensorgram are reduced, as would be expected given the smaller scattering cross-section of the polystyrene nanobeads. Of particular interest in these examples, are events B1 and B2. B1  again corresponds to adsorption of a nanobead at 	a position in which the near field is weak and is thus not detected in the Pearson sensorgram, however event B2 corresponds to adhesion of a particle in close proximity to another previously bound particle. As such, B2 cannot be resolved in the fluorescence image due to the limited spatial resolution of the microscope, albeit it can be seen if cumulative binding images are again calculated due to the slight change in the fluorescence intensity (see Supplementary Movie 2). For all other events indicated, there is a clear correspondence between the two modes of analysis (see also Supplementary Movie 3).

\section{Conclusion}
 Whilst SPR sensors have proven to be a powerful platform in the biomedical sciences, losses in the metallic substrate limit the quality of plasmonic resonances and therefore sensing has correspondingly been restricted to bulk properties. To address this problem, we have proposed using SPR in random metallic substrates which promote generation of speckled leakage radiation. In particular, we have demonstrated that adsorption of single nanoparticles to a random nanoisland substrate can produce detectable decorrelation of the leakage speckle. Such improvements in the detection limit result from the sensitivity of speckle patterns to small changes in the scattering microstructures. Moreover, given that modification of  the speckle pattern arises from interference of the field scattered from the analyte particle and a background speckle, the proposed system exploits interferometric detection to benefit from stronger signal modulations and more robust noise performance. Individual sorption events were observed in the speckle decorrelation sensorgram for both 50 nm radius GNPs and polystyrene nanobeads and verified by means of bright field and fluorescence imaging of the substrate surface respectively.  In summary, our work demonstrates that single particle detection is possible on an SPR platform with minimal system modification. Label-free detection of GNPs and dielectric beads presented in this work relied on physisorption, however, we envision our platform could also be used for specific detection of proteins, viruses  and other small biomolecules through appropriate functionalisation of the sensor surface \cite{Qian2019,ZhangNatMethod2020}. Moreover, given the sensitivity to individual particle motion, our system could be used to monitor the dynamics of molecular machines on the sensor surface or to probe analyte kinetics within a trapping potential \cite{Berk2020}.

	\begin{acknowledgments}
	J.B. and M.R.F. were funded by the Engineering and Physical Sciences Research
	Council (EPSRC) (1992728) and the Royal Society (UF150335) respectively. H.L. and D.K. were supported by the National Research Foundation of Korea (NRF-2019R1A4A1025958).
\end{acknowledgments}

\end{document}